\documentclass{JHEP3}
\usepackage{epsfig}
\title{Nonperturbative definition of  the pole mass and short distance expansion of the heavy quark potential in QCD}
\author{G.\ Grunberg\\
        Centre de Physique Th\'eorique de l' Ecole   Polytechnique (CNRS UMR C7644),\\
        91128 Palaiseau Cedex, France\\
        E-mail: \email{georges.grunberg@pascal.cpht.polytechnique.fr}}
\abstract{We show that the ${\cal O}(\Lambda)$ ambiguity in the pole mass can be fixed in a natural way by
introducing a modified nonperturbative V-scheme momentum space coupling $\tilde{\alpha}_{V}(q)$ where the
confining contributions have been subtracted out. The method used is in the spirit of the infrared finite coupling
approach to power corrections, and gives a nonperturbative definition of the `potential subtracted' mass.  The
short distance expansion of the static potential is derived, taking into account an hypothetical short distance linear term.
The magnitude of the standard OPE contributions are estimated in quenched QCD, based on results of L\"uscher and Weisz. It
is observed that the expansion is not yet reliable  at the shortest distances presently measured on the lattice.} 

\preprint{CPHT/RR 035.0703}
\begin{document}

\section{Introduction}
Historically, the pole mass $M$ and the heavy quark potential $V(r)$ were among the first quantities where
renormalons \cite{Beneke1} have been discussed in a physical context in QCD. Latter, the connection of the ${\cal
O}(\Lambda)$ ambiguity in the  pole mass \cite {Bigi, Beneke-Braun} with a corresponding ambiguity in the coordinate
space potential
\cite{Ligeti} was pointed out. It was observed \cite{Beneke2,Hoang}  that  the leading renormalon contribution
cancels in the total static energy
$E_{static}=2 M+V(r)$, a physical quantity which should be free of ambiguities. This cancellation is a non-trivial
finding. Indeed, one might have expected that the pole mass and the static potential should be separately well
defined: for instance, in the Schr\"odinger equation, the quark mass normalizes the kinetic energy. Furthermore,
although the potential appears to be nonperturbatively defined only up to an arbitrary constant (in particular only
the force is the quantity free of ambiguity in lattice calculations), it is difficult to maintain the view that the
arbitrary normalization of $V(r)$ implies an arbitrary normalization of $M$, which nevertheless would follow from the
non-ambiguity of the static energy if there were no independent way to fix the normalization of either the mass or
the potential. In this paper I suggest that there   is in fact a natural way to define unambiguously the
pole mass at the nonperturbative level (at least as far as the {\em leading} renormalon ambiguity is concerned) even
in a confining theory like QCD, by properly subtracting out the confining contributions to the self-energy, hence to
fix also the `constant term' in the  potential.  In Sec.\ 2, the definition of the
${\cal O}(\Lambda)$ term in the pole mass is given, in term of a properly defined nonperturbative momentum space V-scheme
coupling
$\tilde{\alpha}_{V}(q)$. The method used is in the spirit of the infrared (IR) finite coupling
approach to power corrections \cite{D-W}. In Sec.\ 3, theoretical 
constraints on
$\tilde{\alpha}_{V}(q)$ are reviewed. In Sec.\ 4 the short distance expansion of $V(r)$ is derived, including the effect
of an hypothetical linear short distance term, and the standard IR power corrections are estimated on theoretical ground. It
is shown that present lattice data are not available at distances short enough for a reliable short distance  analysis to be
performed yet.

\section{The nonperturbative pole mass}
To define the pole mass, one has to fix its well-known renormalon
ambiguity \cite{Bigi,Beneke-Braun}. I start from the result \cite{Beneke2,Hoang} that  the leading 
IR contribution $\delta M_{PT\vert IR}$ to the perturbative pole mass
$M_{PT}$ (when expressed in term of a short distance mass like $m\equiv m_{\overline{MS}}$), is related (presumably
to all orders of perturbation theory \cite{Beneke2}) to the leading long distance contribution $\delta V_{PT\vert
IR}$ to the perturbative coordinate space  potential
$V_{PT}$ by the relation

\begin{equation}\delta M_{PT\vert IR}(\mu_{f})=-{1\over 2}\delta V_{PT\vert
IR}(\mu_{f})\label{eq:renM-V1},\end{equation}
where

\begin{equation}\delta V_{PT\vert IR}(\mu_{f})=\int_{\vert\vec
{q}\vert<\mu_{f}}{d^{3}\vec{q}\over(2\pi)^{3}} \tilde{V}_{PT}(q)\label{eq:dVir-pt}.\end{equation}
$\tilde{V}_{PT}(q)$ is the  momentum space perturbative potential, related to $V_{PT}(r)$ by Fourier
transformation 
 
\begin{equation} V_{PT}(r)=\int{d^{3}\vec{q}\over(2\pi)^{3}}\ exp(i\vec{q}.\vec{r})
\tilde{V}_{PT}(q)\label{eq:V},\end{equation}
 and $\mu_{f}$ is an IR factorization scale. Defining to all orders of
perturbation theory a  momentum space  potential effective coupling
$\alpha_{V\vert PT}(q)$ by 

\begin{equation}\tilde{V}_{PT}(q)\equiv -4\pi C_{F}{\alpha_{V\vert PT}(q)\over
q^{2}}\label{eq:apot-pt},\end{equation}
eq.(\ref {eq:renM-V1}) can be rewritten as

\begin{equation}\delta M_{PT\vert IR}(\mu_{f})={C_{F}\over\pi}\int_{0}^{\mu_{f}}dq\
\alpha_{V\vert PT}(q)\label{eq:renM-V2}.\end{equation}
The right hand side of eq.(\ref{eq:renM-V2}) is presumably ill-defined, since it involves an integration over the IR
Landau singularity thought to be present in $\alpha_{V\vert PT}(q)$, and represents (taking $\mu_{f}\sim \Lambda$)
the 
${\cal O}(\Lambda)$ ambiguity in the pole mass. To solve this problem, one would be tempted, in analogy with the IR
finite coupling approach to power corrections \cite{D-W}, to replace the perturbative effective
coupling $\alpha_{V\vert PT}(q)$ inside the integral in eq.(\ref{eq:renM-V2}) by the corresponding {\em
nonperturbative} coupling $\alpha_{V}(q)$ defined by
\begin{equation}\tilde{V}(q)\equiv -4\pi C_{F}{\alpha_{V}(q)\over
q^{2}}\label{eq:apot-np1},\end{equation}
where this time $\tilde{V}(q)$ is the Fourier transform of the full nonperturbative potential $V(r)$:

\begin{equation} V(r)=\int{d^{3}\vec{q}\over(2\pi)^{3}}\ exp(i\vec{q}.\vec{r})
\tilde{V}(q)\label{eq:V-np}.\end{equation}
However, in
a confining theory, $\tilde{V}(q)$ either does not exist (e.g. if $V(r)\sim B\log r+C$ for $r\rightarrow\infty$), or
is anyway  too singular at small
$q$ (reflecting the singular large distance behavior of $V(r)$), making the integral in eq.(\ref{eq:renM-V2}) (with the
nonperturbative $\alpha_{V}(q)$) divergent at $q=0$. For instance, in the case of a linearly raising potential $V(r)={\cal
O}(r)$  for
$r\rightarrow\infty$, one gets $\alpha_{V}(q)={\cal O}(1/q^{2})$ for $q\rightarrow 0$. This observation
suggests one should first subtract out the confining long-distance part of the potential to define a suitable
nonperturbative coupling $\alpha_{V}(q)$. To this end, the following procedure appears the most natural one:
expand the potential around $r=\infty$, and subtract from $V(r)$ the first few leading terms in this expansion
(including an eventual constant term) which do not vanish for $r\rightarrow\infty$. There is by construction only a 
{\em finite} number of such terms. Let us call their sum
$V_{conf}(r)$. Then we have

\begin{equation}V(r)=V_{conf}(r)+\delta V(r)\label{eq:Vconf1},\end{equation}
which, assuming the large $r$ expansion can actually be performed, {\em uniquely} defines $\delta V(r)$, such that $\delta
V(r)\rightarrow 0$ both for
$r\rightarrow 0$ (from asymptotic freedom) {\em and} for
$r\rightarrow\infty$. It is clear that $\delta V(r)$ now admits a standard Fourier representation

\begin{equation} \delta V(r)=\int{d^{3}\vec{q}\over(2\pi)^{3}}\ exp(i\vec{q}.\vec{r})\
\delta\tilde{V}(q)\label{eq:dV},\end{equation}
and one can define the new {\em nonperturbative} coupling $\tilde{\alpha}_{V}(q)$ by

\begin{equation}\delta\tilde{V}(q)\equiv -4\pi C_{F}{\tilde{\alpha}_{V}(q)\over
q^{2}}\label{eq:apot-np2}.\end{equation} 
One should note that the perturbative part of these quantities are preserved, namely $\delta
V_{PT}(r)\equiv V_{PT}(r)$ and $\delta\tilde{V}_{PT}(q)\equiv\tilde{V}_{PT}(q)$, since $\delta V$ differs from $V$
by the $V_{conf}(r)$ term, which, viewed from short distances, appears as a finite sum of 
nonperturbative power-like  corrections, invisible order by order in perturbation theory. Indeed, the terms occurring  in
perturbation theory should scale as
$1/r$, hence vanish for $r\rightarrow\infty$, which excludes them from $V_{conf}(r)$.
Thus $\tilde{\alpha}_{V\vert PT}(q)=\alpha_{V\vert PT}(q)$ is the same as in eq.(\ref{eq:apot-pt}), i.e.
$\tilde{\alpha}_{V}$ and $\alpha_{V}$ have identical perturbative expansions. 

 As an example, consider the potential in quenched
QCD (this is actually the only case where the analytic form of the $r\rightarrow\infty$  expansion is known in low
orders). Theoretical expectations  give the long distance expansion for
$r\rightarrow\infty$

\begin{equation}V(r)\simeq Kr+C-{\pi\over 12}{1\over r}+{\cal O}({1\over r^{2}}),\label{eq:V-as}\end{equation} 
Although the ${\cal O}(1/r)$ term is not a rigorous result of QCD, since it  has been derived within an effective bosonic
string theory
\cite{Luscher}, it has been numerically confirmed \cite{L-W} in high precision lattice simulations. We shall therefore
assume that eq.(\ref{eq:V-as}) gives the correct large distance behavior of the static potential. It follows that
 
\begin{equation}V_{conf}(r)=Kr+C\label{eq:Vconf3},\end{equation}
 and one defines

\begin{equation}V(r)\equiv Kr+C+\delta V(r).\label{eq:Vconf2}\end{equation}
In this case, the couplings $\alpha_{V}(q)$ (if it can be defined nonperturbatively, i.e. if $C=0$  as previously  noted)  
and
$\tilde{\alpha}_{V}(q)$ just differ by a $1/q^2$ term, arising from the Fourier transform of the
$Kr$ piece. 

The prescription for the {\em nonperturbative} definition of
the pole mass now  reads as follows. Introduce the `potential subtracted' mass \cite{Beneke2}

\begin{equation}M_{PS}(\mu_{f})=M_{PT}-\delta M_{PT\vert IR}(\mu_{f}),\label{eq:pot-mass}\end{equation}
and define the nonperturbative IR contribution to the pole mass by

\begin{equation}\delta M_{IR}(\mu_{f})=-{1\over 2}\delta V_{IR}(\mu_{f}),\label{eq:dM-ir}\end{equation}
where

\begin{equation}\delta V_{IR}(\mu_{f})=\int_{\vert\vec
{q}\vert<\mu_{f}}{d^{3}\vec{q}\over(2\pi)^{3}}\ \delta \tilde{V}(q)\label{eq:dVir},\end{equation}
which yields
\begin{equation}\delta M_{IR}(\mu_{f})={C_{F}\over\pi}\int_{0}^{\mu_{f}}dq\
\tilde{\alpha}_{V}(q)\label{eq:V-ir2},\end{equation}
in complete analogy with eq.(\ref{eq:renM-V1}),(\ref{eq:dVir-pt}) and (\ref{eq:renM-V2}). Then the 
pole mass is given by 

\begin{equation}M=M_{PS}(\mu_{f})+\delta M_{IR}(\mu_{f})+...,\label{eq:polemass}\end{equation}
where the dots represent non-leading ${\cal O}(1/m)$ IR  contributions from higher order renormalons, and the
$\mu_{f}$ dependence approximatively cancels between the two terms on the right hand side. The interpretation of
 the prescription eq.(\ref{eq:polemass}) is transparent: it says one should remove from $M_{PT}$ its ambiguous IR
 part $\delta M_{PT\vert IR}(\mu_{f})$, as suggested in \cite{Beneke2}, and substitute for it the corresponding
nonperturbative (and non-ambiguous) IR contribution $\delta M_{IR}(\mu_{f})$. One should note the similarity between
eq.(\ref{eq:polemass}) and the corresponding expressions in the IR finite coupling approach to power corrections
\cite{D-W}. In the present context, however, the nonperturbative coupling is unambiguously identified.
With the pole mass well-defined, the constant term $C$ in the large distance expansion of the potential
(eq.(\ref{eq:V-as})) is in turn fixed, since the corresponding constant term in the large
distance expansion of $E_{static}(r)$, which should be  unambiguous and calculable, is $2M + C$. 

\section{Constraints on the nonperturbative $\tilde{\alpha}_{V}(q)$}

Eq.(\ref{eq:V-as}) and (\ref{eq:Vconf2}) yield $\delta V(r)\sim
-{\pi\over 12}{1\over r}$ for $r\rightarrow\infty$, hence $\delta \tilde{V}(q)\sim -{\pi^{2}\over 3}{1\over q^{2}}$
for $q\rightarrow 0$, which yields 

\begin{equation}C_{F}\ \tilde{\alpha}_{V}(q=0)={\pi\over 12}\label{eq:aIR},\end{equation}
i.e. $\tilde{\alpha}_{V}(q=0)\simeq 0.196$, a rather {\em small} IR fixed point value. Substituting this value as a rough
estimate  of $\tilde{\alpha}_{V}(q)$ in the integrand of eq.(\ref{eq:V-ir2}) gives

\begin{equation}\delta M_{IR}(\mu_{f})\simeq {C_{F}\over\pi}\tilde{\alpha}_{V}(q=0)\
\mu_{f}={1\over 12}\ \mu_{f}\label{eq:V-ir3},\end{equation}
which represents a correction of about $100\ MeV$ for the range of $\mu_{f}$ quoted in \cite{Beneke2} for b-quarks.

A more refined estimate is obtained by inputting the information about the ${\cal O}({1\over r^{2}})$ term in
eq.(\ref{eq:V-as}), which was obtained in \cite{L-W} from a fit to high precision large $r$ lattice data and yields for
$r\rightarrow\infty$

\begin{equation}\delta V(r)\simeq -{\pi\over 12}{1\over r}-{\pi\over 12} {b\over
r^{2}}\label{eq:V-as1}\end{equation}
with $b\simeq 0.04 fm$, hence  for $q\rightarrow 0$

\begin{equation}\delta \tilde{V}(q)\simeq -{\pi^{2}\over 3}{1\over q^{2}}- b\ {\pi^{3}\over 6}{1\over
q}\label{eq:Vtilde-ir}\end{equation}
and

\begin{equation}C_{F}\ \tilde{\alpha}_{V}(q)\simeq{\pi\over 12}\left(1+ b\ {\pi\over 2}\
q\right)\label{eq:aIR1}.\end{equation}
Note that, since $b>0$, $\tilde{\alpha}_{V}(q)$ {\em increases} from its IR value as $q$ increases, hence
must be  non-monotonous  in the IR region, since asymptotic freedom implies it should  ultimately decrease to $0$ at large
$q$. At
$\mu_{f}=1.2\ GeV$, the second term in the parenthesis in eq.(\ref{eq:aIR1}) represents a correction of about $40\%$ to the
IR value. Substituting eq.(\ref{eq:aIR1}) in the integrand of eq.(\ref{eq:V-ir2}) gives

\begin{equation}\delta M_{IR}(\mu_{f})\simeq {1\over 12}\ \mu_{f}\left(1+ b\ {\pi\over 4}\
\mu_{f}\right)\label{eq:V-ir3},\end{equation}
which yields $\delta M_{IR}(\mu_{f})\simeq 120MeV$ for $\mu_{f}=1.2\ GeV$.

\section{Short distance expansion of the heavy quark potential}
 In this section I show that, barring  constant terms, the short distance expansion of the heavy quark potential can be
obtained directly\footnote{I assume  the nonperturbative Fourier transform eq.(\ref{eq:V-np}) exists at
least in a formal sense, in particular that a long distance $B\log r+C$ contribution  is not present, as previously
observed.}  
 from eq.(\ref{eq:V-np}), despite the  singular behavior of $\tilde{V}(q)$ at
small $q$. Introducing again the factorization scale $\mu_{f}$, eq.(\ref{eq:V-np}) can be written as

\begin{equation}  V(r)=-{2\ C_{F}\over\pi}\left[\int_{0}^{\mu_{f}}dq
\left({\sin qr\over qr}\right)\alpha_{V}(q)+\int_{\mu_{f}}^{\infty}dq
\left({\sin qr\over qr}\right)\alpha_{V}(q)\right]\label{eq:V1}.\end{equation}
At short distances, we can expand the $\sin qr$ factor in the low momentum integral, which gives the IR power
corrections. Making the further  assumption that $\alpha_{V}(q)$ has no large power corrections at large $q$ and
may be well approximated by its perturbative part
$\alpha_{V\vert PT}(q)$ above $\mu_{f}$ 

\begin{equation} \alpha_{V}(q)\simeq\alpha_{V\vert PT}(q)\label{eq:alphatildeV-sd0}\end{equation}
(this assumption will be modified below, eq.(\ref{eq:alphatildeV-sd1})), one ends
up with the
$r\rightarrow 0$ expansion

\begin{equation}   V(r)\simeq V_{PT}(r,\mu_{f})-{2\ C_{F}\over\pi}\left[\int_{0}^{\mu_{f}}dq\
\alpha_{V}(q)-{r^{2}\over 6}\int_{0}^{\mu_{f}}dq\
q^{2}\alpha_{V}(q)+{\cal O}(r^{4})\right]\label{eq:V-sd1},\end{equation}
where

\begin{equation} V_{PT}(r,\mu_{f})=-{2\ C_{F}\over\pi}\left[\int_{\mu_{f}}^{\infty}dq
\left({\sin qr\over qr}\right)\alpha_{V\vert PT}(q)\right]\label{eq:V-pt-sub}\end{equation}
is the  IR subtracted perturbative potential \cite{Beneke2}. The normalization of the standard ${\cal O}(r^{0})$ and
${\cal O}(r^{2})$ renormalon-related power corrections in eq.(\ref{eq:V-sd1}) is thus given\footnote{The
corresponding expressions in term of non-local operators can be found in the effective field theory framework of
 \cite{Brambilla}.} by
low-energy moments of
$\alpha_{V}(q)$. Note that the ${\cal O}(r^{0})$ term  is actually infinite, as expected from the
divergent IR  behavior of $\alpha_{V}(q)$. In particular in quenched QCD eq.(\ref{eq:aIR1}) implies for $q^2\rightarrow 0$

\begin{equation}C_{F}\ \alpha_{V}(q)\sim {2\ K\over q^{2}}+{\pi\over 12}\left(1+ b\ {\pi\over 2}\
q\right)\label{eq:aV-IR}.\end{equation}
But since the ${\cal O}(r^{0})$ term contributes only an overall normalization constant to the potential, which in this
section is left arbitrary, one can drop it out. On the other hand, the
${\cal O}(r^{2})$ and higher order r-dependent contributions are {\em finite}. In particular, using eq.(\ref{eq:aV-IR}) as a
rough approximation to
$\alpha_{V}(q)$ in the range $0<q<\mu_{f}$ one obtains in quenched QCD for
$r\rightarrow 0$ (ignoring any constant term)

\begin{equation}  V(r)\simeq V_{PT}(r,\mu_{f})+\left({2K\over3\pi}\mu_{f}+{1\over 108}\mu_{f}^{3}+{b\pi\over
288}\mu_{f}^{4}\right) r^{2}+{\cal O}(r^{4})\label{eq:V-sd2}.\end{equation}

Let us now modify the previously mentioned assumption, in order to deal with the possibility that a ${\cal
O}(1/q^{2})$  power correction is actually present in $\alpha_{V}(q)$. Such a correction has been first suggested in
\cite {Zakharov} as a consequence of new physics  related to confinement, leading to a ${\cal O}(r)$
linear correction to the potential at short distances, of the same size (and sign) as the standard long distance
correction related to the string tension.  It
should be noted however that  a short distance linear piece may have a more conventional (although still non
perturbative) {\em infrared} origin, as indicated  by the position of  the leading IR  renormalon present
in $\tilde{V}(q)$, which also suggests \cite{Beneke2} the presence of a
${\cal O}(1/q^{2})$ correction. Let us thus assume that for $q^2\rightarrow \infty$

\begin{equation} \alpha_{V}(q)\simeq\alpha_{V\vert PT}(q)+{2K_{0}\over C_{F}}{1\over
q^2}\label{eq:alphatildeV-sd1}\end{equation}
with $K_{0}\neq K$ in general. To deal with this correction, one can use the general method of \cite{Gru-power}, or
more conveniently,  introduce a new coupling
$\tilde{\alpha}_{V}(q)$ (different in general from the one in section 2, see  below), related to the original 
$\alpha_{V}(q)$ by

\begin{equation}C_{F}\alpha_{V}(q) \equiv C_{F}\tilde{\alpha}_{V}(q)+ {2K_{0}\over
q^2}\label{eq:alphaV-alphatildeV},\end{equation}
such that the {\em redefined} coupling $\tilde{\alpha}_{V}(q)$ is essentially given by its perturbative part (which
coincides with that of $\alpha_{V}(q)$) at large
$q^2$

\begin{equation} \tilde{\alpha}_{V}(q)\simeq \alpha_{V\vert PT}(q)\label{eq:alphaV-sd1}\end{equation}
 with no substantial power corrections. Thus from eq.(\ref{eq:alphaV-alphatildeV})

\begin{equation}\tilde{V}(q)=-{8\pi K_{0}\over q^{4}} -4\pi C_{F}{\tilde{\alpha}_{V}(q)\over
q^{2}} \label{eq:apot-np1},\end{equation}
and, upon taking the Fourier transform

\begin{equation}V(r)=K_{0}\ r+\delta V(r)\label{eq:deltaV2},\end{equation}
where  $\delta V(r)$ is given by eq.(\ref{eq:dV}) and (\ref{eq:apot-np2}), but with 
$\tilde{\alpha}_{V}(q)$ now defined by eq.(\ref{eq:alphaV-alphatildeV}). Note that for  $K_{0}=K$  this definition
coincides  with that of section 2 (assuming $C=0$, see the comment after eq.(\ref{eq:Vconf2})). Thus, introducing  a
factorization scale $\mu_{f}$ as in eq.(\ref{eq:V1}) we have  

\begin{equation} \delta V(r)=-{2\ C_{F}\over\pi}\left[\int_{0}^{\mu_{f}}dq
\left({\sin qr\over qr}\right)\tilde{\alpha}_{V}(q)+\int_{\mu_{f}}^{\infty}dq
\left({\sin qr\over qr}\right)\tilde{\alpha}_{V}(q)\right]\label{eq:dV1}.\end{equation}
Since $ \tilde{\alpha}_{V}(q)$ has no large
power corrections, it can be approximated by its perturbative part  $\alpha_{V\vert PT}(q)$
above some scale
$\mu_{f}$, and  one deduces the short distance expansion

\begin{equation} \delta  V(r)\simeq V_{PT}(r,\mu_{f})-{2\ C_{F}\over\pi}\left[\int_{0}^{\mu_{f}}dq\
\tilde{\alpha}_{V}(q)-{r^{2}\over 6}\int_{0}^{\mu_{f}}dq\
q^{2}\tilde{\alpha}_{V}(q)+{\cal O}(r^{4})\right]\label{eq:dV-sd1}.\end{equation}
 From eq.(\ref{eq:aV-IR}) and (\ref{eq:alphaV-alphatildeV}) we get for
$q^2\rightarrow 0$

\begin{equation}C_{F}\ \tilde{\alpha}_{V}(q)\sim 2\ (K-K_{0}){1\over q^{2}}+{\pi\over 12}\left(1+ b\ {\pi\over 2}\
q\right)\label{eq:atildeV-IR}.\end{equation}
Thus, dropping again the (infinite) ${\cal O}(r^{0})$ term, and using eq.(\ref{eq:atildeV-IR}) for $q< \mu_{f}$, we
obtain  

\begin{equation} \delta V(r)\simeq V_{PT}(r,\mu_{f})+\left({2(K-K_{0})\over3\pi}\mu_{f}+{1\over 108}\mu_{f}^{3}+{b\pi\over
288}\mu_{f}^{4}\right)
r^{2}+{\cal O}(r^{4})\label{eq:dV-sd2},\end{equation}
hence from eq.(\ref{eq:deltaV2})

\begin{equation}  V(r)\simeq V_{PT}(r,\mu_{f})+K_{0}\ r+\left({2(K-K_{0})\over3\pi}\mu_{f}+{1\over 108}\mu_{f}^{3}+{b\pi\over
288}\mu_{f}^{4}\right)
r^{2}+{\cal O}(r^{4})\label{eq:V-sd3}\end{equation}
which of course agrees with eq.(\ref{eq:V-sd2}) for $K_{0}=0$. The correlation between the coefficient of the ${\cal
O}(r)$ correction (which is
$\mu_{f}$ independent) and that of the standard OPE 
${\cal O}(r^{2})$   correction should be noted. For $K_{0}\neq 0$, we get a neat
derivation of the well-known statement \cite{Zakharov} that the appearance of a linear short distance term in $V(r)$ is
equivalent to the presence of a ${\cal O}(1/q^2)$ correction in the standard $\alpha_{V}(q)$ coupling. Moreover, for
$K_{0}=K$, one obtains the straightforward, but interesting, result that the appearance  of the  linear short distance 
 term  is equivalent to the statement that  the {\em modified} coupling $\tilde{\alpha}_{V}(q)$ of section 2 (rather
then
$\alpha_{V}(q)$)  has no 
${\cal O}(1/q^2)$ corrections.

One might attempt  an analysis of the lattice short distance data of \cite{Sommer} based on
eq.(\ref{eq:deltaV2}) and (\ref{eq:dV-sd1}).
$V_{PT}(r,\mu_{f})$  could be evaluated from eq.(\ref{eq:V-pt-sub}) by solving  the known \cite{Schroder} 3-loop
renormalization group equation for
$\alpha_{V\vert PT}(q)$ and performing  the integral, similar to the single dressed gluon `renormalon integral' (with IR
cut-off) in
\cite{Gru-power, Gar-Gru}, while the power corrections should be fitted.
Unfortunately, one finds that the perturbative expansion
of the V-scheme coupling beta function is not reliable at values of
$\mu_{f}$ small enough that the low momentum integral in eq.(\ref{eq:dV1}) can be meaningfully expanded and parametrized in
term of a few power correction terms, even at the shortest values of $r$  presently measured on the lattice. Thus no
reliable fit of the power corrections can be performed yet. It should be noted that in the present approach  standard IR
power corrections appear from an OPE-like separation of long and short distances in the Fourier transform of the momentum
space potential, and their presence is mandatory. This is to be contrasted with the result of \cite{Sommer}, where no power
corrections were needed\footnote{The alternative analysis of \cite{Pineda} also finds no room for power corrections.} if the
potential is predicted in term of the renormalization group equation of the position space effective charge $\alpha_{F}$ 
associated
\cite{Gru-force} to the force $F(r)={dV\over dr}= C_F {\alpha_{F}(1/r)\over r^2}$. However, the implicit definition\footnote{The convergence of the  expansion of the  $\alpha_{F}$ beta
function is only slightly better
\cite{Sommer} than that of the momentum space $\alpha_{V}$ beta function, which presumably  makes  a quantitative analysis
of the power corrections difficult also in the scheme of \cite{Sommer}.} of
the power corrections in the later case is different, and does not make use of a momentum space IR cutoff to 
 separate  long from short distances.

\section{Conclusion}
We have shown that it is possible to fix in a natural way the ${\cal O} (\Lambda)$ renormalon ambiguity in the pole
mass, thus giving a {\em nonperturbative} definition of the pole mass in QCD at this level of accuracy, which represents
a natural nonperturbative extension of the `potential subtracted' mass, in the spirit of the IR finite coupling approach to
power corrections. This definition is an optimal one, in the sense the prescription is to remove from the heavy quark
potential contribution to the  self-energy  those terms  and {\em only} those one (the confining ones contained in
$V_{conf}(r)$) which would give a meaningless (infinite) result for the pole mass. For instance,   one should {\em not}
remove from
$\delta V(r)$ the ${\cal O}(1/r)$ `L\"uscher term' to include it in $V_{conf}(r)$ (see eq.(\ref{eq:V-as}))
(which, moreover, would make the modified IR finite V-scheme coupling
$\tilde{\alpha}_{V}(q)$ non-asymptotically free!). The applications of the proposed
mass definition are similar to those of the `potential subtracted' mass, to which it provides the leading power correction,
allowing an accurate relation to the standard $\overline {MS}$ mass,  but it can be used consistently   with non-perturbative
extensions of the Coulomb static potential (such as implied by phenomenological potential models or the potential
determined on the lattice).  The remaining challenge is to fix the  ${\cal O} (\Lambda^2/m)$ ambiguities in the
pole mass arising from higher order renormalons. 

We  have also discussed the OPE like analysis of the short distance
potential. The
magnitude of the standard OPE contributions have been estimated from eq.(\ref{eq:aV-IR}).
However,  the resulting short distance expansion is unreliable at the lowest values of $r$
measured so far on the lattice, due to the poor convergence of perturbation theory for the momentum space V-scheme coupling
beta function.

\acknowledgments I thank N. Brambilla, Yu.L. Dokshitzer, G. Marchesini and A. Vairo for useful  discussions, and A.
Pineda for a correspondence. I also wish to thank the referee for helpful comments.

%\newpage


\begin{thebibliography}{9}

\bibitem{Beneke1} M. Beneke, {\em Phys. Rept.} {\bf 317} (1999) 1, and references therein.

\bibitem{Bigi} I.I. Bigi, M.A. Shifman, N.G. Uraltsev and A.I. Vainshtein, {\em Phys.Rev.} {\bf D50} (1994) 2234.


\bibitem{Beneke-Braun} M. Beneke and V.M. Braun, {\em Nucl. Phys.} {\bf B426} (1994) 301; M. Beneke, {\em Phys.
Lett.} {\bf B344} (1995) 341.

\bibitem{Ligeti} U. Aglietti and Z. Ligeti, {\em Phys. Lett.} {\bf B364} (1995) 75.

\bibitem{Beneke2} M. Beneke, {\em Phys. Lett.} {\bf B434} (1998) 115.  

\bibitem{Hoang} A.H. Hoang, M.C. Smith, T. Stelzer and S. Willenbrock, {\em Phys. Rev.} {\bf D59} (1999) 114014;
A. Pineda, {\em Heavy quarkonium and nonrelativistic effective theories}, PhD thesis, Barcelona (1998).
  

\bibitem{D-W} Yu.L. Dokshitzer and B.R. Webber, {\em Phys. Lett.} {\bf B352} (1995) 451; Yu.L. Dokshitzer, G.
Marchesini and B.R. Webber, {\em Nucl. Phys.} {\bf B469} (1996) 93. See also M. Neubert, {\em Phys. Rev.} {\bf
D51} (1995) 5924.
 
\bibitem{Luscher} M. L\"uscher, K. Zymanzik and P. Weisz, {\em Nucl. Phys.} {\bf B173} (1980) 365; M. L\"uscher,
{\em Nucl. Phys.} {\bf B180} (1981) 317.

\bibitem{L-W} M. L\"uscher and P. Weisz, {\em JHEP} {\bf 0207} (2002) 049.




\bibitem{Brambilla} N. Brambilla, A. Pineda, J. Soto and A. Vairo, {\em Nucl. Phys.} {\bf B566} (2000) 275. 



\bibitem{Zakharov} M.N. Chernodub, F.V.
Gubarev, M.I. Polikarpov and V.I. Zakharov, {\em Phys. Lett.} {\bf B475} (2000) 303 , and references therein.


\bibitem{Gru-power} G. Grunberg, {\em JHEP} {\bf 9811} (1998) 006.



\bibitem{Sommer} S. Necco and R. Sommer, {\em Phys. Lett.} {\bf B523} (2001) 135, and references therein.

\bibitem{Schroder} M. Peter, {\em Phys. Rev. Lett.} {\bf 78} (1997) 602; Y. Schr\"oder, {\em Phys. Lett.} {\bf B447} (1999)
321.

\bibitem{Gar-Gru} E. Gardi and G. Grunberg, {\em JHEP} {\bf 9911} (1999) 016.

\bibitem{Pineda}  A. Pineda, {\em J.Phys.} {\bf G29} (2003) 371.

\bibitem{Gru-force} G. Grunberg, {\em Phys. Rev.} {\bf D40} (1989) 680.








 

\end{thebibliography}
\end{document}